\documentclass[12pt]{article}
\usepackage{amssymb,amsfonts,epsfig}
\newcommand{\be}{\begin{equation}}
\newcommand{\ee}{\end{equation}}
\newcommand{\bea}{\begin{eqnarray}}
\newcommand{\eea}{\end{eqnarray}}   
\newcommand{\vp}{\varphi}
\newcommand{\bref}[1]{(\ref{#1})}

\def\ie{{\it i.e. }}

\def\t{\theta}

\def\del3{\delta^{(3)}}

\def\undos{{1 \over 2}}

\def\e{\epsilon}
\def\otaula{\begin{tabular}}
\def\ctaula{\end{tabular}}
\def\espai{\;\;\;\;\;\;}
\def\zespai{\;\;\;\;}

\def\G{\Gamma}

\def\gym{g_{YM}}

\def\a{\alpha}

\def\avall{\vspace{1cm}}

\def\ka{K\"ahler }
\def\2ka{K\"ahler}
\def\et{\tilde{e}}

\def\T{\Theta}
\def\CN{$\mathcal{N}$}
\def\tH{\tilde{H}}
\def\th{\tilde{h}}

\def\oz{\overline{z}}

\def\MR{\mathbb{R}}

\begin{document}

\begin{titlepage}

\setcounter{page}{0}
\begin{flushright}
UB-ECM-PF-02/28 \\
\end{flushright}

\vspace{5mm}
\begin{center}
{\Large {\bf Commutative and noncommutative \CN=2 
SYM in 2+1 from wrapped D6-branes
}}
\vspace{10mm}

{\large Jan Brugu\'es, Joaquim Gomis, Toni Mateos, 
Toni Ram\'{\i}rez} \\
\vspace{5mm}
{\em Departament ECM, Facultat de F{\'\i}sica, \\
Institut de Fisica d'Altes Energies and\\
CER for Astrophysics, Particle Physics and Cosmology, \\
Universitat de Barcelona \\
Diagonal 647, E-08028 Barcelona, Spain }\\
\vspace{5mm}
\end{center}
\vspace{20mm}

\centerline{{\bf{Abstract}}}

\vspace{5mm}

We give the supergravity duals
of commutative and noncommutative non-abelian gauge theories 
with \CN=2 in 2+1 dimensions.
The moduli space on the Coulomb branch of these theories
is studied using supergravity.

\avall
\avall
\avall
\avall
\avall

\avall

Proceedings of the RTN Workshop, {\it ``The quantum structure
of spacetime and the geometric nature of fundamental interactions},
Leuven, September 2002. Based on a talk given by J.G. and including 
new material.

\vfill{
 \hrule width 5.cm
\vskip 2.mm
{\small
\noindent E-mail: jan,gomis,tonim,tonir@ecm.ub.es \\
}}

\end{titlepage}
\newpage
\section{Introduction and Conclusions}

In the last two years, an extension of the AdS/CFT \cite{Maldacena:1997re}
correspondence
to gauge field theories with less than maximal supersymmetry has been
achieved via wrapped branes,
and many supergravity duals have been constructed 
\cite{Maldacena:2000mw}-\cite{quim-jo}.
Gauged supergravities provided a useful tool to construct such configurations
since they geometrically implement the twisting  
\cite{Bershadsky:1995qy} of the field theories
in a natural way. 
On the other hand,
supergravity duals of noncommutative (NC) theories with
maximal supersymmetry had also been constructed 
\cite{Hashimoto:1999ut} by turning on a background $B$-field. See 
\cite{Larsson:2001wt}
for a 
summary of all flat Dp-brane solutions.

In \cite{Brugues:2002ff},  both ideas were joined
and the dual of a $U(N)$ NC field theory 
\CN=2 in 2+1 dimensions was constructed. 
Unlike most cases of wrapped branes obtained before,
the noncommutative problem had
to be analysed directly in 11d supergravity,
since it was shown that reducing the background to 8d gauged
supergravity would have destroyed all supersymmetry.
In this work, we will review the original construction
of these backgrounds, 
emphasising more on the phenomenon of supersymmetry
without supersymmetry \cite{duff}.

We will also illustrate some of the field theory
physics that can be extracted from the supergravity side
by obtaining the moduli space on the Coulomb branch. In the
commutative case, this corrects the calculation originally
performed in \cite{quim-jo}, where supersymmetry had actually
been lost in the reduction from eleven to ten dimensions.
We obtain a two-dimensional moduli space
which is \ka (as demanded by supersymmetry, and in
agreement with \cite{jaume-gauntlett}) and which,
in fact, looks very much like a resummation of all the
perturbative contributions of the quantum field theory.
This is not in contradiction with the known fact that
non-abelian gauge theories with \CN=2 and without
matter fields develop a non-perturbative potential that completely
lifts the Coulomb branch \cite{wittenaffleck}. Such effects are not
expected to be seen in the gravity side, since they are exponentially
suppressed at large $N$.

We reanalysed the same problem in the noncommutative case and we found
exactly the same results, {\it i.e.} the moduli space and its metric
coincide with the commutative ones, in agreement with the discussion
in \cite{polchinski}. Other non-perturbative properties of a NC SYM
theory with \CN=1 in 3+1 dimensions were studied in \cite{Mateos:2002rx}.

The paper is organised as follows. In section 2 we review how the
noncommutative duals were constructed in \cite{quim-jo}, simplifying
the discussion of which compactifications are compatible with
supersymmetry. In section 3 we calculate the moduli spaces of
the commutative and the noncommutative theories, and we compare them
with the field theory results.

\section{Supergravity duals}

Since the appearance of \cite{Maldacena:2000mw}, gauged supergravities
became a systematic tool to obtain supergravity solitons 
describing the near horizon of wrapped branes.
Essentially, one chooses a domain-wall kind of ansatz (with the
expected isometries of the configuration) in the appropriate gauged supergravity.
Then, imposing that some supersymmetry is preserved automatically leads
to a system of first order BPS differential equations.
The solutions are then lifted to ten or eleven dimensions
giving the corresponding sugra dual. 

In \cite{quim-jo}, such method was used to obtain the dual of an \CN=2 field
theory
in 2+1 dimensions by wrapping D6-branes on \ka four-cycles inside
Calabi-Yau three-folds.
We used 8d gauged supergravity \cite{salam}, 
which is  a compactification of the
eleven-dimensional one on an $SU(2)$ manifold.
When trying to perform a noncommutative deformation on such solution, one
could naively try to repeat the mentioned process, but now extending the 8d ansatz to
incorporate the modifications. To illustrate why this method would fail
we will briefly analyse the much simpler case of the NC deformation of
a flat (not wrapped) D6-brane.
Such a configuration was already known, and its lift to 11d
\cite{Larsson:2001wt} is
\be \label{nearh} ds^2_{11}= h^{1/3} \left[dx_{0,4}^2 +
 h^{-{1}} dx_{5,6}^2 +
dy^2+ {y^2\over 4} \left( d\Omega_{(2)}^2 +
h^{-1} [ d\psi+\cos\theta d\phi ]^2 \right)\right] \ee \be
A_{[3]}= -\T{ N \gym^2 \over H \, h}  \, dx^5\wedge
dx^6 \wedge \left( d\psi +\cos\theta d\phi\right) \zespai\zespai
h(y)=1+\T^2H^{-1} \zespai\zespai H(y)={4 N^2 \gym^4 \over y^2}.
\ee
A way to analyse which of the possible reductions would preserve
supersymmetry is to compute the associated Killing spinors in the appropriate
vielbein base. For example, to reduce back to 10d along the $U(1)$ isometries
generated by $\partial_{\psi}$ or by $\partial_{\phi}$,
we need to take the first ten vielbeins to
be independent of $\psi$ or $\phi$ respectively. A good choice for both could be
\be
e^a=h^{1/6}\,dx^a\, , \,\, a=0,..,4 \espai\espai
e^{i}=h^{-1/3}\, dx^{i}\, , \,\, i=5,6 \espai\espai
 e^7=h^{1/6} \, dy,
\ee
\be
\label{vi} e^8= {y\over 2} h^{1/6} \, \et^1 \espai\espai
e^9= {y\over 2} h^{1/6}  \, \et^2 \espai\espai e^T=
 {y\over 2} h^{-1/3}
\, \et^3, \ee with $\et^i$ the typical vielbeins of a round $S^3$ 
\be \label{e-base}
\et^1=d \t \espai\espai \et^2=\sin\t d\phi \espai\espai
\et^3=d\psi+\cos\t d\phi \, . 
\ee 
On the other hand, if we want to reduce to 8d gauged sugra, we need
to use the $SU(2)$ left-invariant one-forms for the $S^3$, so we have
to replace \bref{e-base} by
\be w^1=-\cos\psi d\theta - \sin\theta \sin\psi d\phi \zespai\zespai
w^2=-\sin\psi d\theta + \sin\theta \cos\psi d\phi \zespai\zespai
w^3=-d\psi-\cos\theta d\phi. \label{vi3} \ee 
We will call \bref{e-base} the $e$-base and \bref{vi3} the $w$-base.
Since the relation between them is just a local Lorentz rotation, the Killing
spinors in both bases are simply related by its spin $\undos$ representation.
Explicit calculation yields the following 16 spinors
\be \label{spinsol2} \e_{{\rm (e-base)}}
=h^{{1\over 12}}(y)   e^{{\a(y)\over 2} \G_{78}}
e^{{\theta\over 2} \G_{78}}
e^{{\phi\over 2}\G_{89}} \e_0, \espai\espai 
\ee\be
\label{rotation} \e_{{\rm (w-base)}}\,=\,e^{-{\psi\over 2} \G_{89}}
e^{{\pi\over 2} \G_{T8}}\,\e_{{\rm (e-base)}}\,=\, \G_{T8}
e^{{\psi\over 2} \G_{89}}\,
\e_{{\rm (e-base)}}.
\ee
with
\be
\G_{0123456}\e_0=-{\e}_0 \espai\espai \cos\a=h^{-\undos} \espai\espai
\sin\a=\T(Hh)^{-\undos}
\ee

Based on the fact that both compactifications assume that the spinors 
are independent of the coordinates of the compact manifold, 
it is immediate to see that: ({\it i}) Supersymmetry is preserved under
compactification to IIA along $\psi$ but completely destroyed under
compactification along $\phi$; ({\it ii}) Supersymmetry is
completely destroyed under compactification to 8d gauged sugra.

These conclusions were explicitely checked. Indeed, since all fields
but the Killing spinors fit in the corresponding reduction ansatze, the
{\it wrong} compactifications produced good solutions of the equations
of motion, despite being non-supersymmetric.

With this experience in mind, it was natural to construct the NC deformation
of the wrapped D6 directly in eleven dimensions \cite{Brugues:2002ff}.
We solved the BPS equations for our ansatz and obtained the Killing spinors. 
Just like before, by analysing them in the correct reduction base, we could
show that:
({\it i}) A reduction to 8d sugra would not have been supersymmetric;
({\it ii}) A reduction to IIA along $\phi$ would preserve supersymmetry,
but a reduction along $\psi$ would destroy it.


Finally, the correct reduction to IIA (along $\phi$) gave the true supergravity
dual of the NC \CN=2 gauge theory in 2+1 dimensions. After introducing
the factors of the number $N$ of D6-branes, and the string coupling constant
$g_s$, the background is\footnote{The
various definitions appearing here are: $U(r)={3r^4+8l^2r^2+6l^4 \over
6(r^2+l^2)^2};\, f(r,\t)=\sin^2\t+U \cos^2\t;\, m(r,\t)=(U^{-1}+\cot^2\t)^{-1};\, 
B_{[1]}=d\psi+\cos\t_1 d\phi_1+\cos\t_2 d\phi_2 ;\, \tH={4\over r^2f};\,
\th=1+\T^2\tH^{-1}$. For simplicity, the four-cycle will be taken 
to be $S^2\times S^2$ in all this work.}
\be \label{nc-metrica}
ds^2_{IIA}=\left(g_s N \tH\right) ^{-\undos}
 \left[ -dx_0^2+ \th^{-1} dx_{1,2}^2 \right]+
\left(g_s N \right)^{\undos} \left[{3\over 2}(r^2+l^2)ds^2_{cycle}+U^{-1}dr^2+
{r^2 \over 4} [ d\t^2 + m B_{[1]}^2]\right]
\ee\be
e^{4\Phi/3}=g_s^{1\over 3}N^{-1}
\th^{-{2\over 3}} \tH ^{-1} \espai\espai
B_{[2]}=-{\T\over g_s N}{1 \over \tH \th}\, dx^1\wedge dx^2 \ee\be
C_{[1]}=-N\, Uf^{-1}\cos\t\,B_{[1]} \espai\espai
C_{[3]}=-{\T\over g_s\tH \th}Uf^{-1}\cos\t\,\,\, dx^1\wedge dx^2 \wedge B_{[1]}.
\ee
and describes a non-threshold bound state of D6 and D4 branes, all of them
wrapping the \ka four-cycle, and with the D4 spread in the flat part of the D6.
This can be best described with the following array

\begin{center}
\be
\begin{array}{c | c c c c c c c c c c}
{\rm IIA} &x^0 & x^1 & x^2 & \t_1 & \t_2 & \phi_2 & \phi_1 & r &\t &
\psi \\ \hline
{\rm D6}       &-&-&-&-&-&-&-&&& \\
{\rm D4}       &-& & &-&-&-&-&&& \\
B_{[2]}  & &-&-& & & & &&& \\
\end{array}\ee
\end{center}
Note that the metric has cohomogeneity two. The function $\tH$ depends,
after a suitable change of variables, on the transverse coordinate inside
the Calabi-Yau three-fold, and on the transverse one to the D6-branes
and the Calabi-Yau.

The supergravity approximation is valid where the curvature and
the dilaton remain small. In this case, these restrictions imply
\be \label{validity}
{1\over (g_s N)^{1/6}}\, \ll \, r \ll \,{N^{1/4} \over g_s^{1/12}} \,\,\,.
\ee

\section{Analysis of the moduli spaces}

In this section we use the constructed supergravity duals of 
the \CN=2 theories to extract some physics. In particular, we will
obtain and discuss the moduli spaces on the Coulomb branch in
both the commutative and the noncommutative cases, and we will 
compare them with the expected results from the field theory side.

\subsection{The commutative moduli space from supergravity}

Ley us first concentrate on the commutative theory. The supergravity background
is just obtained by sending $\T\rightarrow 0$ in \bref{nc-metrica}, and is dual
to an $SU(N)$ gauge theory with \CN=2 in 2+1 dimensions, without any matter
multiplet.

We will  analyse the Coulomb branch of this theory by giving
a non-zero vacuum expectation value to the scalars in a $U(1)$ subgroup
of $SU(N)$. As is well known, this is easily implemented in the supergravity
side by pulling one of the $N$ D6-branes away from the others. 
The $U(1)$ degrees of freedom on the probe brane can be effectively described
by the DBI action, where the rest of the branes are substituted by the 
background that they create
\be
\mu_6^{-1}S=-
\int_{\Sigma_7}
d^7\xi \,\, e^{-\Phi}
\sqrt{-\det[G+B_{[2]}+2\pi\a'F_{[2]}]}+  
\int_{\Sigma_7}
[\exp(2\pi\a'F+B)\wedge \oplus_n C_{[n]}] \label{dbi}
\ee
Here $\mu_6^{-1}= {(2\pi)}^6\a'^{7/2}$, $F_{[2]}=dV_{[1]}$ is the worldvolume
Abelian field-strength, $B_{[2]}$ is the NS two-form,
$C_{[n]}$ the RR $n$-forms, and all fields are understood
to be pulled-back to the seven-dimensional worldvolume
$\Sigma_7=\MR^{1,2}\times S^2 \times S^2$.

If we want to break the gauge group without breaking supersymmetry, we
must make sure that no potential is generated.
So the first thing to look at is the vacuum configuration of the probe brane.
With this purpose, we take the static gauge where the first seven space-time
coordinates are identified with the worldvolume ones, and all the rest,
\ie $\{r,\t,\psi\}$, are taken to be constant. In this way, only the
potential is left in the DBI action. It is not possible to give a
closed analytic
expression for it but, numerically, it is easy to see that it vanishes
only at $\t=0$ and $\t=\pi$, independently of $r$ and $\psi$.

We therefore locate the probe brane at such values of $\t$ and look at
the low energy effective action for its massless degrees of freedom.
This is accomplished by allowing $\{r,\psi\}$ and the worldvolume field-strength
$F_{[2]}$ to slowly depend on the worldvolume coordinates, so that only
the terms quadratic in their derivatives a kept in the expansion of the DBI action.
Indeed, in the limit in which the four-cycle is taken to be small, one can 
simply consider excitations about the flat non-compact part of the worldvolume.
Both locus $\t=0,\pi$ give the same effective action:
\be
-S_{probe}=\int d^3x
\left[ a^2 N (g_sN)^{3/2}C^2(r) r^2 \,(\partial r)^2 \, \, +\, \,
{1\over g_s N^2}{1\over 4 C^2(r)} \, (\partial y)^2 \right]
\ee
where $a^2=2 \pi^2\mu_6$, $C(r)={1\over 4}(r^2+l^2)$
and $y$ is the compact scalar of period $2\pi$ that
one obtains after dualising the gauge field $V_{[1]}$.

The moduli space is therefore two-dimensional and, after glueing the two
locus at the origin, it turns out to have the topology of a cylinder.
The metric is just
\be \label{sugra-metric}
ds^2
=a^2 N (g_sN)^{3/2} C^2(r)\,
 r^2 \,dr^2 \, \, +\, \, {1\over g_s N^2}{1\over 4 C^2(r)}  \, dy^2 \,\,=
\,\, d\rho^2+{4 a  \over a g_s N^2 l^4 + 16 (g_s N^3)^{1/4} \,\rho}\, dy^2
\ee
In the last step we redefined the radial coordinate 
$\rho={a g^{3/ 4}N^{5/ 4}\over 16}\, r^2\left(r^2+2l^2\right)$ 
in order to put the metric in a more standard form. It is easy to prove
that this metric is \ka by explicitely constructing the \ka potential.
In order to do so, first define complex coordinates
\be
z=y+i\chi(r) \espai\espai \chi(r):={a\over 48}N(g_sN)^{5/4}
\left(r^6+3l^2 r^4+3l^4 r^2\right)
\ee
One can then show that $ds^2_{{\rm moduli}}\,=\,2\, g_{z\oz}\, dz \otimes d\oz$
with $g_{z\oz}=\partial_z \partial_{\oz} \vp$ and
\be
\vp
 = {a^{2/3}\over 128}g_s^{1/2}N^{5/6}\left( a g_s^{3/4} N^{5/4}l^6
+{48 \over g_s^{1/2}N} \,\, {z-\oz \over 2i} \right)^{4\over 3} 
={a^2 N (g_sN)^{3/2}\over 128}(r^2+l^2)^4
\ee
The fact that $\vp$ is a real function completes the proof that the metric 
is \2ka (see \cite{jaume-gauntlett} for similar results using different
branes) .

\subsection{Comparison with the field theory results}

We shall now compare the results obtained using supergravity with the
ones that are known from the field theory.
The first immediate comment is that in the absence of matter multiplets,
instantons of non-abelian gauge theories with \CN=2 in 2+1 dimensions
develop a superpotential that completely lifts the Coulomb branch
\cite{wittenaffleck}.
This is not in contradiction with our result, since this contributions
are exponentially suppressed with $N$, so they are not expected to
be visible in the supergravity side.

On the other hand, \CN=2 supersymmetry implies that the moduli space
must be a 2d \ka manifold, as we have seen from supergravity. Furthermore,
it typically has the topology of a cylinder, with the compact direction 
coming from the dualised scalar, and the non-compact one coming
from the vacuum expectation value of the other scalar in the multiplet.
The one loop corrected metric for an $SU(2)$ theory \cite{hori} looks like
\be \label{1loop}
ds^2={1\over 4}\left({{1\over e^2}-{2\over r}}\right) dr^2+
\left({{1\over e^2}-{2\over r}}\right)^{-1}dy^2
\ee
and it is valid for $r \gg 2e^2$. Asymptotically, it tends to 
the classical prediction which, after generalising to $SU(N)$,
is just a flat cylinder with metric
\be
ds^2={1\over 4 e^2} dr^2+{e^2 \over N} dy^2
\ee

In order to compare these metrics with our supergravity result \bref{sugra-metric}
we shall perform a change of variables in \bref{1loop} so that the metric is in 
the standard form $ds^2=d\rho^2+f(\rho) dy^2$. Unfortunately, the change of
variables is not expressible in terms of elementary functions. Anyway, we can
solve numerically for $f(\rho)$ and plot the two moduli spaces, as we have
done in figure 1.
\begin{figure}
\begin{center} \label{radi}
\includegraphics[width=6cm,height=4cm]{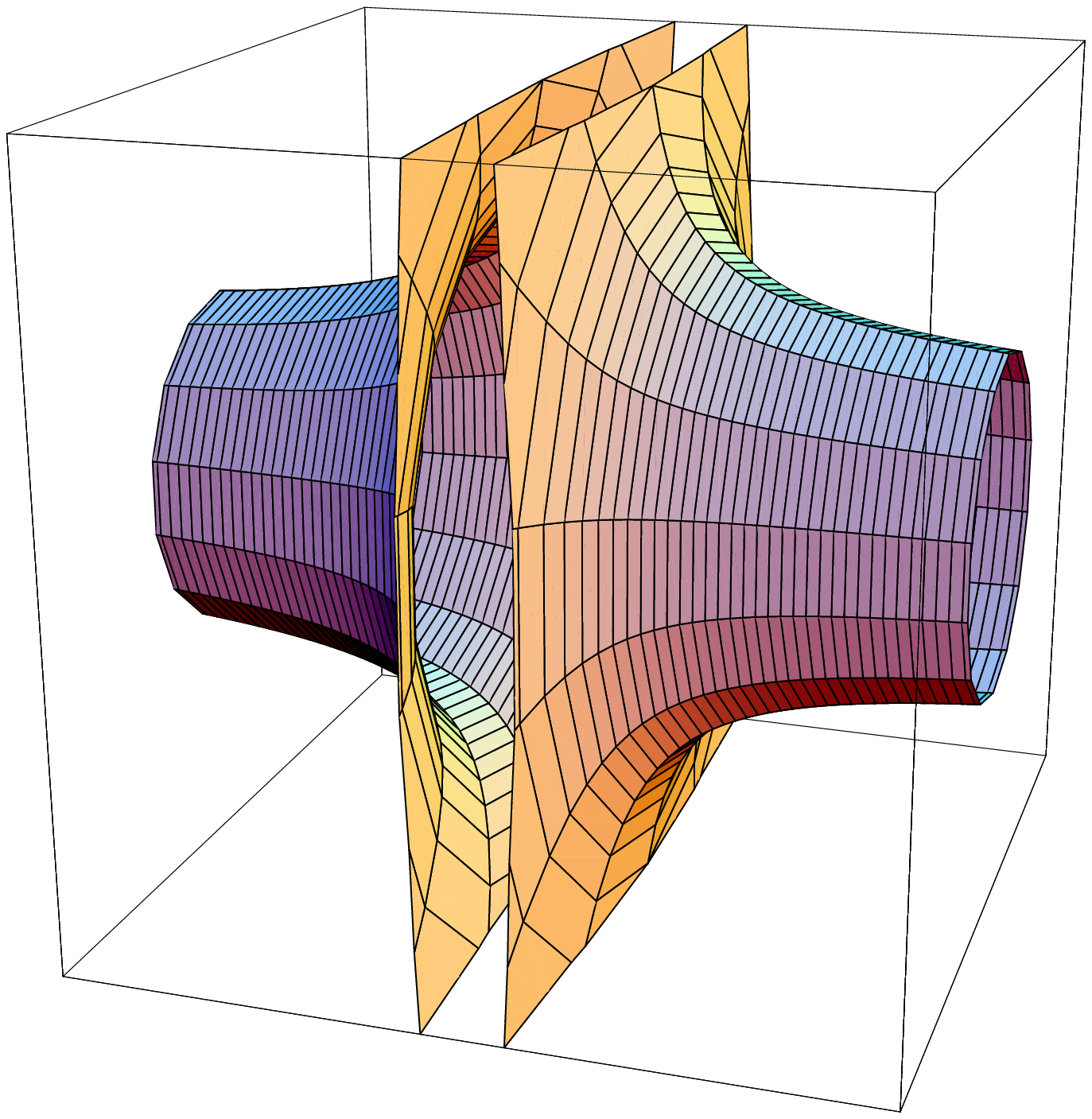} $\espai\espai$
\includegraphics[width=6cm,height=4cm]{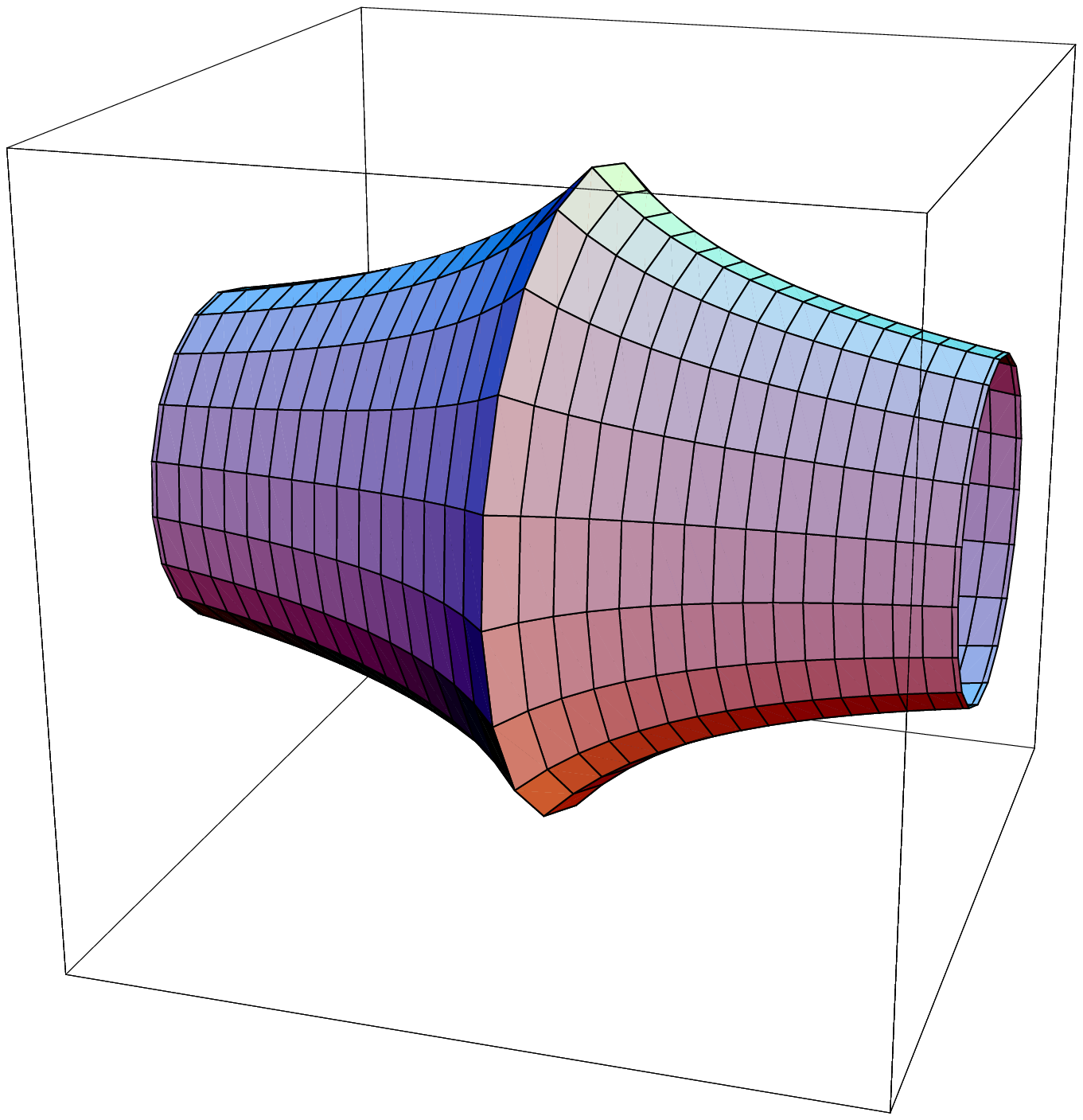}
\end{center}
\caption{Moduli spaces from 1-loop field theory (left) and supergravity (right).}
\end{figure}
The plot on the left shows the one-loop corrected moduli space predicted by
field theory calculations. At very large values of the non-compact scalar,
it  tends to flat cylinder with radius proportional to $|e|/\sqrt{N}$.
As this $vev$ decreases higher loop corrections are needed. In particular,
the one loop calculation diverges at $r=2 e^2$.

On the other hand, the figure on the right shows the moduli space predicted
by supergravity. It also tends to a cylinder with vanishing radius at large 
values of the non-compact scalar, so it agrees with the $N \rightarrow \infty$
limit of the field theory. It also smooths the divergence of the one loop
calculation, which could maybe  correspond to a resummation
of infinite loops contributions. Strictly speaking, we see from \bref{validity}
that the supergravity approximation is not valid at $r=0$, where the curvature
of our background blows up. In any case, we can still use it as close to the
origin as needed by taking $g_s N$ large enough.

Finally, we shall make more explicit the
relation between the parameters in supergravity ($g_s$, $N$ and $l$)
and in the field theory ($e$ and $N$).
As usual, the number $N$ of D6-branes is the rank of the gauge group.
On the other hand, in the supergravity side,
a non-zero value for $l$ prevents the radius from diverging
as we approach the origin.
Nevertheless, it is difficult
to make the dictionary more precise. In any case, one can read the
gauge coupling for the $U(1)$ degrees of freedom at a certain
point of the moduli space by identifying the coefficient in front
of the $F^2$ term in the DBI action of probe. The result is
\be
{1 \over g_{U(1)}^2}=4 \pi^2 \mu_6 \, g_s N^2 \, (r^2+l^2)
\ee

\subsection{The noncommutative moduli space}

Exactly the same method used to obtain the moduli space of the commutative
field theory can be used to study the noncommutative one. In this case
we have to use the full NC background \bref{nc-metrica}.
The vacuum of the probe brane is the same. This is
due to the fact that, in the static gauge, and without exciting the fields on
the probe, one can verify that our background satisfies
\be
e^{-\Phi^{\T}}\sqrt{-\det[G^{\T}]}=e^{-\Phi^{\T=0}}\sqrt{-\det[G^{\T=0}]}\,,
\espai\espai\espai
C^{\T}_{[7]}+B^{\T}_{[2]}\wedge C^{\T}_{[5]}=C^{\T=0}_{[7]} 
\ee
In superscripts we indicate whether the field is written at finite or at zero value
for $\T$, and pullbacks are to be understood where needed. 
Since the DBI lagrangian is unchanged, the supersymmetric loci are still at
$\t=0,\pi$.

When we look for the low energy excitations of the fields on the brane, 
we find again the same metric on the moduli space. This time, the relevant
property of the background is
\be
e^{-\Phi^{\T}}\sqrt{-\det[G^{\T}+B_{[2]}^{\T}+2\pi \a ' F_{[2]}]}
+2\pi\a ' F_{[2]}\wedge C^{\T}_{[5]}
= e^{-\Phi^{\T=0}}\sqrt{-\det[G^{\T=0}+2\pi \a ' F_{[2]}]}
\ee
where the equality is to be understood only between the terms which are
quadratic in the derivatives on both sides.

\vskip 6mm

{\it{\bf Acknowledgements}}

We are grateful to 
Jaume Gomis, Jorge Russo, Luca Tagliacozzo and Paul Townsend
for useful discussions. 
This work is partially supported by MCYT FPA, 2001-3598,
 and CIRIT, GC 2001SGR-00065, and HPRN-CT-2000-00131.
T.M. is supported by a grant from the Commissionat per
a la Recerca de la Generalitat de Catalunya. 
J.B. is supported
by a grant from Ministerio de Ciencia y Tecnolog\'{\i}a.


\vskip 4mm









\begin{thebibliography}{99}

\bibitem{Maldacena:1997re}
J.~M.~Maldacena,
Adv.\ Theor.\ Math.\ Phys.\  {\bf 2} (1998) 231
[Int.\ J.\ Theor.\ Phys.\  {\bf 38} (1999) 1113]
[arXiv:hep-th/9711200].
S.~S.~Gubser, I.~R.~Klebanov and A.~M.~Polyakov,
Phys.\ Lett.\ B {\bf 428} (1998) 105
[arXiv:hep-th/9802109].
E.~Witten,
Adv.\ Theor.\ Math.\ Phys.\  {\bf 2} (1998) 253
[arXiv:hep-th/9802150].


\bibitem{Maldacena:2000mw}
J.~M.~Maldacena and C.~Nunez,
Int.\ J.\ Mod.\ Phys.\ A {\bf 16} (2001) 822
[arXiv:hep-th/0007018].


\bibitem{sugrapack}
J.~M.~Maldacena and C.~Nunez,
Phys.\ Rev.\ Lett.\  {\bf 86} (2001) 588
[arXiv:hep-th/0008001].
B.~S.~Acharya, J.~P.~Gauntlett and N.~Kim,
Phys.\ Rev.\ D {\bf 63} (2001) 106003
[arXiv:hep-th/0011190].
H.~Nieder and Y.~Oz,
JHEP {\bf 0103} (2001) 008
[arXiv:hep-th/0011288].
J.~P.~Gauntlett, N.~Kim and D.~Waldram,
Phys.\ Rev.\ D {\bf 63} (2001) 126001
[arXiv:hep-th/0012195].
C.~Nunez, I.~Y.~Park, M.~Schvellinger and T.~A.~Tran,
JHEP {\bf 0104} (2001) 025
[arXiv:hep-th/0103080].
J.~D.~Edelstein and C.~Nunez,
JHEP {\bf 0104} (2001) 028
[arXiv:hep-th/0103167].
M.~Schvellinger and T.~A.~Tran,
JHEP {\bf 0106} (2001) 025
[arXiv:hep-th/0105019].
J.~M.~Maldacena and H.~Nastase,
JHEP {\bf 0109} (2001) 024
[arXiv:hep-th/0105049].
J.~P.~Gauntlett, N.~Kim, S.~Pakis and D.~Waldram,
Phys.\ Rev.\ D {\bf 65} (2002) 026003
[arXiv:hep-th/0105250].
R.~Hernandez,
Phys.\ Lett.\ B {\bf 521} (2001) 371
[arXiv:hep-th/0106055].
J.~P.~Gauntlett, N.~Kim, D.~Martelli and D.~Waldram,
Phys.\ Rev.\ D {\bf 64} (2001) 106008
[arXiv:hep-th/0106117].
F.~Bigazzi, A.~L.~Cotrone and A.~Zaffaroni,
Phys.\ Lett.\ B {\bf 519} (2001) 269
[arXiv:hep-th/0106160].
J.~P.~Gauntlett and N.~Kim,
Phys.\ Rev.\ D {\bf 65} (2002) 086003
[arXiv:hep-th/0109039].
J.~Gomis,
Nucl.\ Phys.\ B {\bf 624} (2002) 181
[arXiv:hep-th/0111060].
G.~Curio, B.~Kors and D.~Lust,
arXiv:hep-th/0111165.
P.~Di Vecchia, H.~Enger, E.~Imeroni and E.~Lozano-Tellechea,
Nucl.\ Phys.\ B {\bf 631} (2002) 95
[arXiv:hep-th/0112126].
R.~Hernandez and K.~Sfetsos,
Phys.\ Lett.\ B {\bf 536} (2002) 294
[arXiv:hep-th/0202135].
J.~P.~Gauntlett, N.~Kim, S.~Pakis and D.~Waldram,
arXiv:hep-th/0202184.
U.~Gursoy, C.~Nunez and M.~Schvellinger,
JHEP {\bf 0206} (2002) 015
[arXiv:hep-th/0203124].
R.~Hernandez and K.~Sfetsos,
arXiv:hep-th/0205099.
P.~Di Vecchia, A.~Lerda and P.~Merlatti,
arXiv:hep-th/0205204.
R.~Apreda, F.~Bigazzi, A.~L.~Cotrone, M.~Petrini and A.~Zaffaroni,
Phys.\ Lett.\ B {\bf 536} (2002) 161
[arXiv:hep-th/0112236].
M.~Naka,
arXiv:hep-th/0206141.
J.~D.~Edelstein, A.~Paredes and A.~V.~Ramallo,
arXiv:hep-th/0207127.
R.~Hernandez and K.~Sfetsos,
arXiv:hep-th/0211130.
J.~D.~Edelstein, A.~Paredes and A.~V.~Ramallo,
arXiv:hep-th/0211203.
J.~D.~Edelstein,
arXiv:hep-th/0211204.
J.~D.~Edelstein, A.~Paredes and A.~V.~Ramallo,
arXiv:hep-th/0212139.



\bibitem{jaume-gauntlett}
J.~Gomis and J.~G.~Russo,
JHEP {\bf 0110} (2001) 028
[arXiv:hep-th/0109177].
J.~P.~Gauntlett, N.~w.~Kim, D.~Martelli and D.~Waldram,
JHEP {\bf 0111} (2001) 018
[arXiv:hep-th/0110034].





\bibitem{quim-jo}
J.~Gomis and T.~Mateos,
Phys.\ Lett.\ B {\bf 524} (2002) 170
[arXiv:hep-th/0108080].




\bibitem{Bershadsky:1995qy}
M.~Bershadsky, C.~Vafa and V.~Sadov,
Nucl.\ Phys.\ B {\bf 463} (1996) 420
[arXiv:hep-th/9511222].





\bibitem{Hashimoto:1999ut}
A.~Hashimoto and N.~Itzhaki,
Phys.\ Lett.\ B {\bf 465} (1999) 142
[arXiv:hep-th/9907166].
J.~M.~Maldacena and J.~G.~Russo,
JHEP {\bf 9909} (1999) 025
[arXiv:hep-th/9908134].
M.~Alishahiha, Y.~Oz and M.~M.~Sheikh-Jabbari,
JHEP {\bf 9911} (1999) 007
[arXiv:hep-th/9909215].
D.~S.~Berman {\it et al.},
JHEP {\bf 0105} (2001) 002
[arXiv:hep-th/0011282].

\bibitem{Larsson:2001wt}
H.~Larsson,
Class.\ Quant.\ Grav.\  {\bf 19} (2002) 2689
[arXiv:hep-th/0105083].


\bibitem{Brugues:2002ff}
J.~Brugues, J.~Gomis, T.~Mateos and T.~Ramirez,
JHEP {\bf 0210} (2002) 016
[arXiv:hep-th/0207091].





\bibitem{duff}
M.~J.~Duff, H.~Lu and C.~N.~Pope,
Phys.\ Lett.\ B {\bf 409} (1997) 136
[arXiv:hep-th/9704186].


\bibitem{wittenaffleck}
I.~Affleck, J.~A.~Harvey and E.~Witten,
Nucl.\ Phys.\ B {\bf 206} (1982) 413.

\bibitem{polchinski}
A.~Buchel, A.~W.~Peet and J.~Polchinski,
Phys.\ Rev.\ D {\bf 63} (2001) 044009
[arXiv:hep-th/0008076].

\bibitem{Mateos:2002rx}
T.~Mateos, J.~M.~Pons and P.~Talavera,
arXiv:hep-th/0209150.


\bibitem{hori}
J.~de Boer, K.~Hori and Y.~Oz,
Nucl.\ Phys.\ B {\bf 500} (1997) 163
[arXiv:hep-th/9703100].
O.~Aharony, A.~Hanany, K.~A.~Intriligator, N.~Seiberg and M.~J.~Strassler,
Nucl.\ Phys.\ B {\bf 499} (1997) 67
[arXiv:hep-th/9703110].

\bibitem{salam}
A.~Salam and E.~Sezgin,
Nucl.\ Phys.\ B {\bf 258} (1985) 284.




\end{thebibliography}
\end{document}